# PRELIMINARY INVESTIGATIONS OF MONTE CARLO SIMULATIONS OF NEUTRON ENERGY AND LET SPECTRA FOR FAST NEUTRON THERAPY FACILITIES

*T.K. Kroc*

*Fermi National Accelerator Laboratory[1], Mail Stop 301, Kirk and Wilson Streets, Batavia, Illinois 60510, USA*

**Abstract:**
No fast neutron therapy facility has been built with optimized beam quality based on a thorough understanding of the neutron spectrum and its resulting biological effectiveness. A study has been initiated to provide the information necessary for such an optimization. Monte Carlo studies will be used to simulate neutron energy spectra and LET spectra. These studies will be bench-marked with data taken at existing fast neutron therapy facilities. Results will also be compared with radiobiological studies to further support beam quality optimization. These simulations, anchored by this data, will then be used to determine what parameters might be optimized to take full advantage of the unique LET properties of fast neutron beams.

This paper will present preliminary work in generating energy and LET spectra for the Fermilab fast neutron therapy facility.

**Introduction:**
The sophistication of current Monte Carlo codes and the computing power of common desktop machines provides the opportunity to investigate the details of the interaction of radiotherapy particles to a degree not possible when neutron therapy was first developed.

This work is the beginning of a thorough investigation into the parameters involved in producing neutron therapy beams. These include the materials and configuration of targets and collimators, beam energy, energy spectra modifications, and beam modulation and shaping. The goal is to optimize these parameters to produce better clinical results and reduce complications.

Monte Carlo codes are now able to tally Linear Energy Transfer (LET). With this we can look back at historic and present facilities to see if varying the parameters mentioned above produce differences in LET which can be correlated with clinical results. We can also look forward to see if we can predict the results of radiobiological investigations. Correlations between these parameters and the response of different tissue types may lead to new optimization algorithms for treatment planning programs.

The results presented here were generated with MCNPX versions 2.6 (1) and 2.7.B (2). However, other persons affiliated with the Neutron Therapy Facility (NTF) at Fermilab

---

are working with GEANT4 (3). Cross checking between the two codes along with comparison with dosimetric data taken in phantoms will increase confidence in the results of this work and its applicability to future plans.

**Initial Simulations:**
The essential components of the target at NTF is a slug of beryllium (2.54 cm diam. × 2.21 cm) followed by a gold foil (2.54 cm diam. × 0.05 cm). It is conventionally thought that the gold foil behind the 49 MeV beryllium target is supposed to range out the remaining energy of the 66 MeV proton beam. Examination of log books from the time the target system was designed reveal the range calculations. However, initial simulations shown here suggest that the calculations did not take range straggling into account.

The early simulations, conducted while learning to operate MCNPX, looked at the simple geometry of just the beryllium and the gold foil. Figures 1a, 1b, and 1c show the flux of protons, neutrons, and photons in the target area. One can clearly see the protons punching through the gold (1a). The maximum neutron flux appears to occur at a shallower depth than 49 MeV (1b). The maximum photon flux occurs where most of the protons range out (1c). The level of detail available in simulations such as these should allow us to fine tune target design for future facilities to maximize the neutron yield and possibly minimize the photon yield. The energy spectrum of the protons emerging from the gold peaks at about 2 MeV and extends almost to 10 MeV as seen in the red curve of figure 2.

To further ensure that the proton punch through was a real effect, the beryllium and gold were segmented into a number of small slices in order to see the evolution of the proton spectrum as it passed through the target assembly. This evolution is seen in Figure 2. Here one can see the broadening of the proton peak resulting from range straggling. The area under the curves appears to remain constant from the initial 66 MeV spike through the white curve. The final curve, in red, is the spectrum of the protons emerging from the gold. Here the area under the last curve is clearly reduced indicating that the gold was almost, but not quite, sufficient to stop all the protons. Also shown is the result of a similar simulation run(4) with GEANT4 (green) which compares very well with the MCPNX results. Comparisons such as this will be used as additional benchmarking to validate the results of the simulations.

In the actual, physical setup, the beryllium and gold foil is placed in a target holder which includes an eighth inch of aluminum downstream of the gold. Subsequent simulations show that this is sufficient to ensure the range out of the protons.

**LET:**
One of the primary goals of these investigations is to determine, as much as possible, the biological effect of neutron therapy beams. The ability to simulate the Linear Energy Transfer (LET) is a big part of this goal. MCNPX has just recently developed the ability to determine the LET of charged particles.

Figure 3, adapted from IAEA TECDOC-992 (5), shows the area of LET where neutrons are presumed to have an advantage over protons and photons. It also identifies the major components that contribute to the high LET component. Neutrons, being uncharged, interact through nuclear processes and produce low energy protons, alphas, and recoil

ions which are high LET.  Figure 4 shows the LET spectra for a number of charged particles from MCNPX runs of 60 MeV (diamonds) and 20 MeV (crosses) mono-energetic neutrons incident on a cube of A-150 plastic. The plot is very preliminary but shows the significant contribution from protons, alphas and to a lesser extent, recoil ions. Comparing the trends between the two energies, one sees that proton LET increases with lower incident neutron energy as one expects. This is also somewhat true for the recoil ions. However, the alpha contribution decreases significantly with the lower neutron energy. Deuterons are not quite a significant contribution at 60 MeV but the trend suggests that they may be at higher energy.

At least two issues remain in the analysis of these results. One is to reconcile the display of the data so that the Monte Carlo data can be understood in the same light as the IAEA plot. The other is to further understand the trends of production of high LET particles such that they can be optimized for the best clinical result.

**Facility Comparisons:**
It is hoped that the ability to perform detailed simulations of LET will allow us to do comparisons between fast neutron therapy facilities. Table 1 shows the other facilities we hope to simulate. Fermilab, Seattle, Detroit, and South Africa are the present, operational, high energy facilities. Understanding details of each of those beams may give insight into their clinical results and may be able to guide their future use. Clatterbridge and UCLA are historical facilities with similar targeting to the existing facilities. The uniqueness of the neutron generation at the Detroit facility presents an intriguing contrast to the other facilities. Hammersmith is where all this started. Its deuteron beam can be compared to Detroit's. These simulations may help in interpreting Hammersmith's initial clinical results.

**Figure 1**

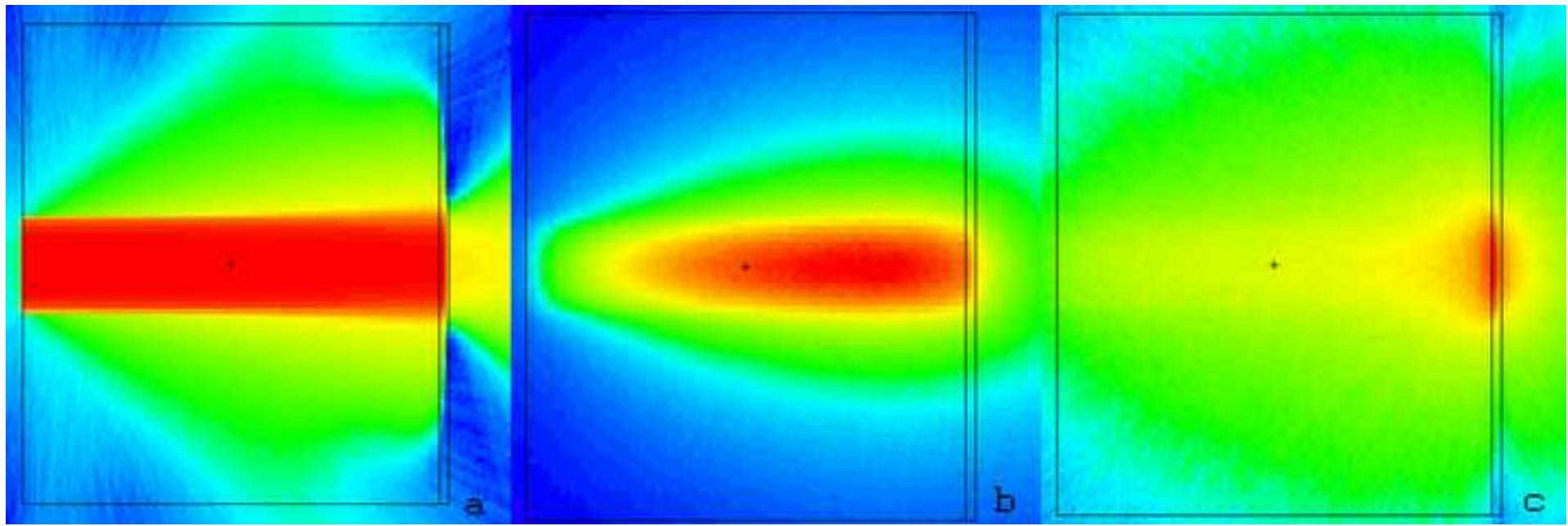

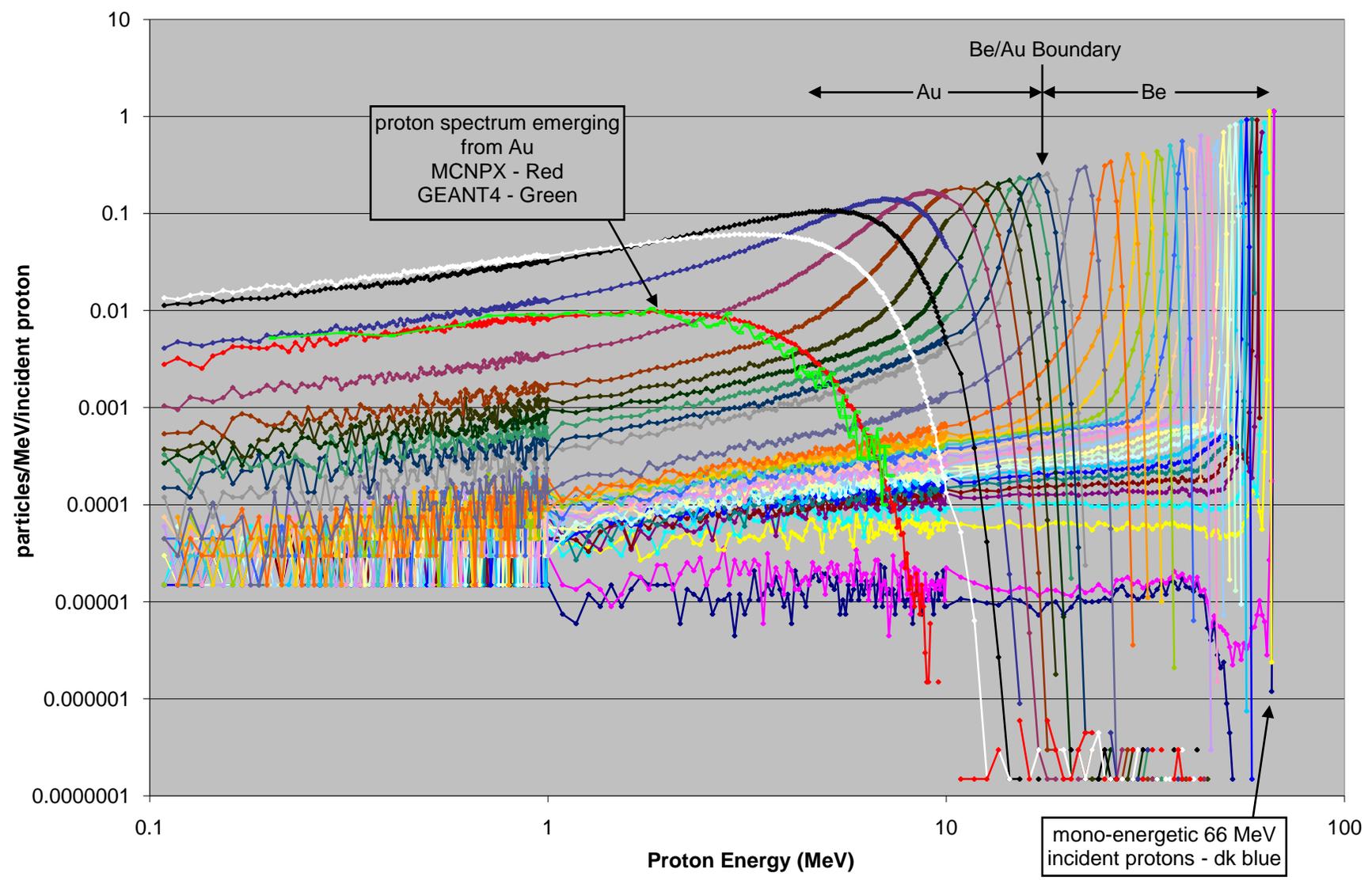

**Figure 3**

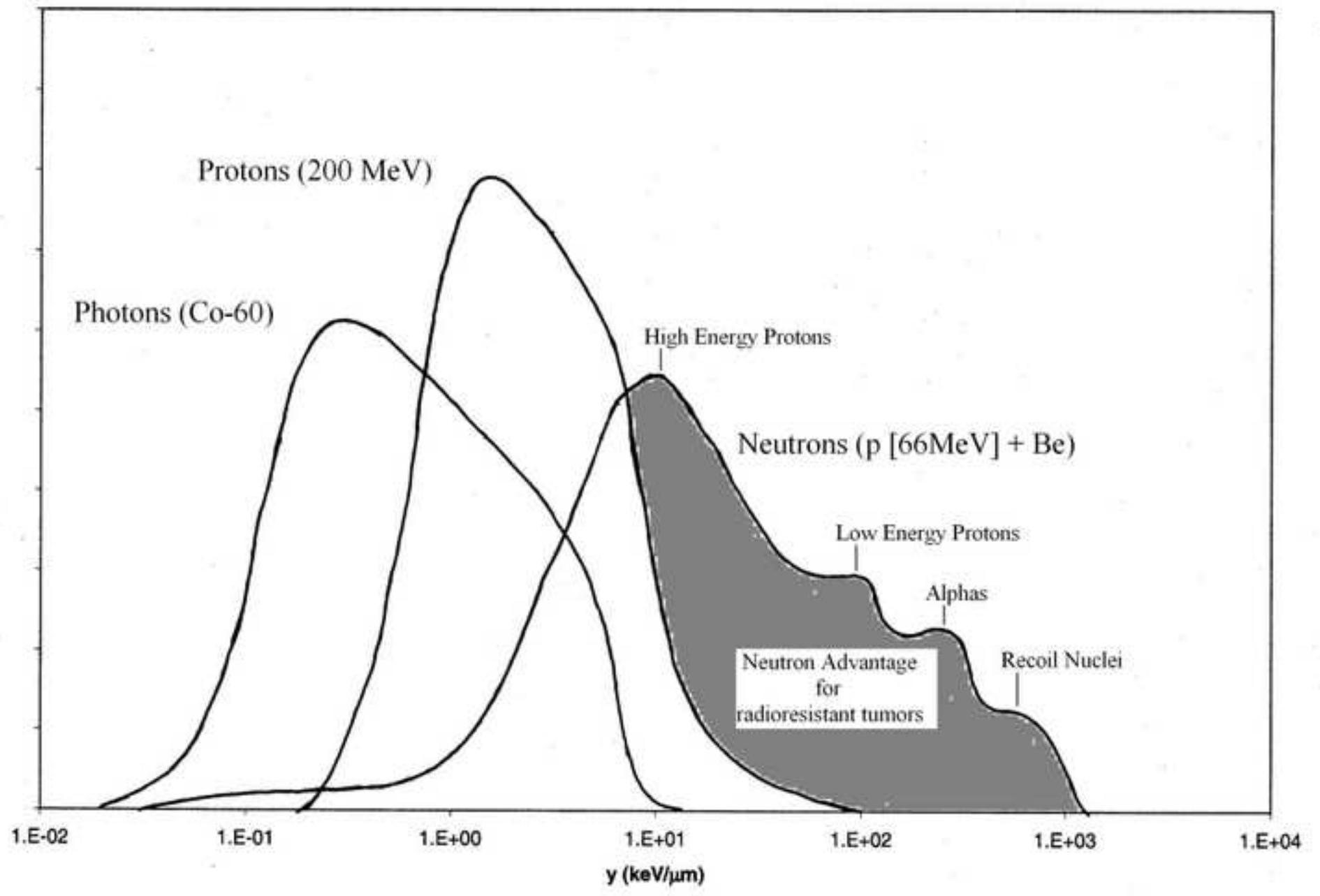

Figure 4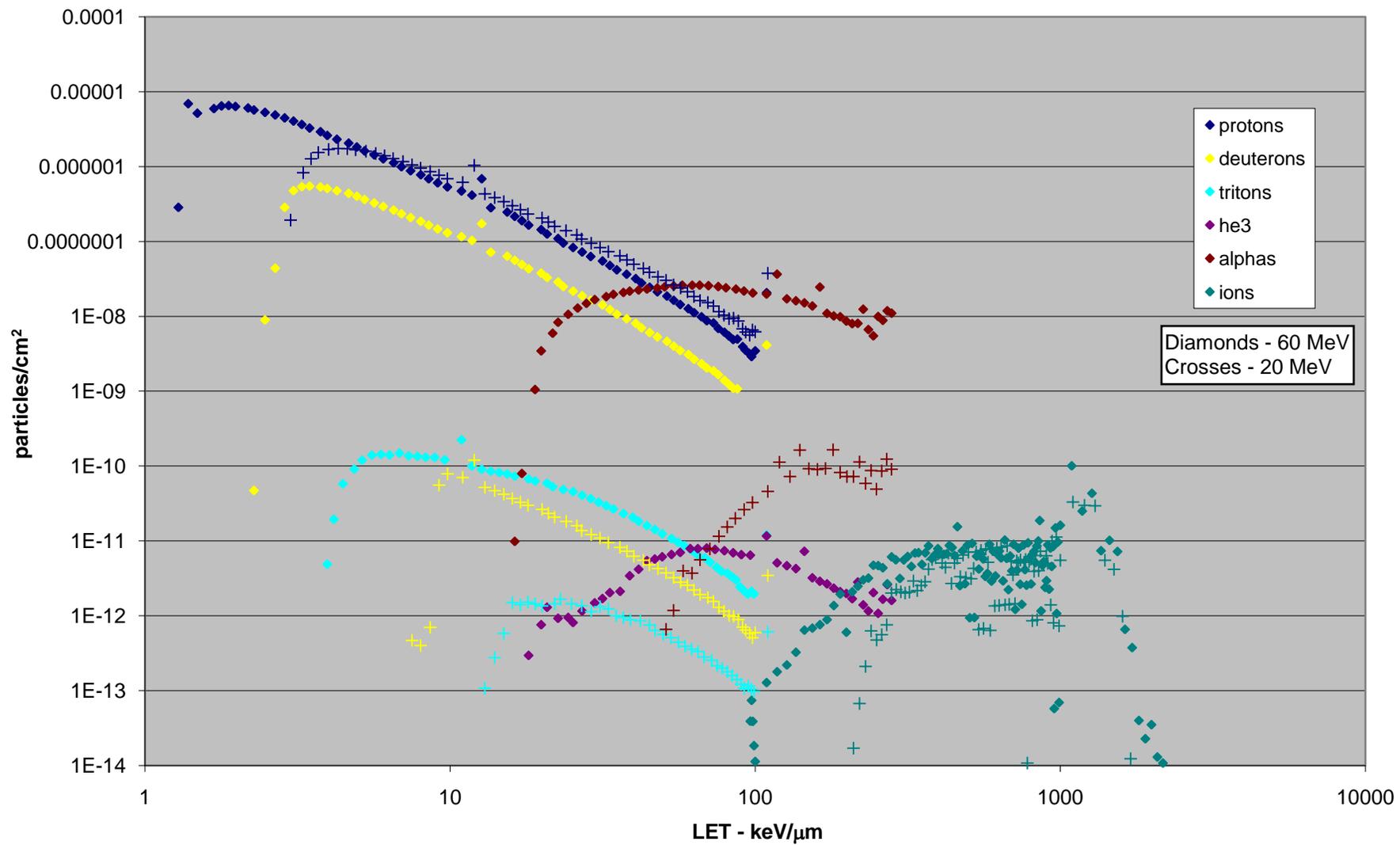

**Table 1**

| Facility | Beam Quality | Target (cm) | Collimation/Comments |
|---|---|---|---|
| Fermilab | p(66) + Be(49) | 2.2 Be, .05 Au, .3 Al | Concrete/Poly or Polyurethane inserts |
| Seattle | p(50) + Be | 1.05 Be, 0.3 Cu, 0.2 $H_2O$, 0.2 PB | Fe + Poly? |
| Detroit | d(50) + Be | Glancing internal | W / Effect of internal target? |
| Clatterbridge | p(62) + Be(36) | 1.8 cm Be | |
| Hammersmith | d(16) + Be | | The place that started it all |
| UCLA | p(46) + Be | 1.0 Be, 0.25 C, 0.375 Cu, 0.125 $H_2O$ | |
| NAC | p(66) + Be(40) | 1.96 Be + ? | Fe, Borated Poly |

Figure Captions

**Figure 1. Particle flux in target.**
Particle flux (protons [a], neutrons [b], photons [c]) in beryllium target (2.54 cm diam. × 2.21 cm) and gold foil (2.54 cm diam. × 0.05 cm). Proton beam enters from left. Color scale of flux (#/cm$^2$) ranges from dark blue (minimum) to red (maximum). Numeric ranges are $1.5\times10^{-5}$ to $6.1\times10^{-1}$ [a], $7.3\times10^{-5}$ to $6.3\times10^{-3}$ [b], $6.3\times10^{-4}$ to $2.1\times10^{-2}$ [c].

**Figure 2. Proton spectrum as a function of penetration into the target.**
Evolution of the energy spectrum of the proton beam as it passes through 34 segments of beryllium and gold target.

**Figure 3. Neutron LET spectrum relative to proton and photon LET spectra.**
Area in gray highlights LET advantage of neutrons with secondary particles noted. Vertical scale is arbitrary but proportional to y·d(y).

**Figure 4. LET spectrum of the flux of secondary particles from 20 and 60 MeV neutrons.**
Secondary particles simulated by MCNPX in a cube with composition of A-150 plastic. Sensitive area for LET tally is a 10 cm cube within the 40 cm A-150 cube. Tally cube center was 10 cm deep along beam axis and centered transversely. Incident beam was either 20 or 60 MeV mono-energetic neutrons.

**Table 1. Proposed facilities for simulation.**
Therapeutic neutron facilities (current and historical) that we propose for comparison of neutron and LET spectra.